\documentclass[conference]{ieeeconf}
\usepackage[bookmarks=true]{hyperref}
\usepackage{cite}
\usepackage{amsmath,amssymb,amsfonts}
\usepackage{algorithmic}
\usepackage{graphicx}
\usepackage{subcaption}
\usepackage{textcomp}
\usepackage{xcolor}
\usepackage{multirow}
\DeclareUnicodeCharacter{2217}{*} 
\def\BibTeX{{\rm B\kern-.05em{\sc i\kern-.025em b}\kern-.08em
    T\kern-.1667em\lower.7ex\hbox{E}\kern-.125emX}}
\begin{document}
\newcommand{\comment}[1]{{\color{red}#1}}
\setlength{\belowcaptionskip}{-10pt}

\newcommand{\refsec}[1]{Section~\ref{sec:#1}}
\newcommand{\refsubsec}[1]{Section~\ref{subsec:#1}}
\newcommand{\reffig}[1]{Fig.~\ref{fig:#1}}
\newcommand{\reftab}[1]{Table~\ref{tab:#1}}
\newcommand{\refeqn}[1]{Equation~\ref{eqn:#1}}

\title{Efficient Computation of Map-scale Continuous Mutual Information on Chip in Real Time}

\author{Keshav Gupta, Peter Zhi Xuan Li, Sertac Karaman, Vivienne Sze
\\Massachusetts Institute of Technology, Cambridge, Massachusetts 02139
\\Emails: \texttt{\{keshav21,peterli,sertac,sze\}@mit.edu}}
\maketitle

\begin{abstract}
Exploration tasks are essential to many emerging robotics applications, ranging from search and rescue to space exploration. The planning problem for exploration requires determining the best locations for future measurements that will enhance the fidelity of the map, for example, by reducing its total entropy. A widely-studied technique involves computing the Mutual Information (MI) between the current map and future measurements, and utilizing this MI metric to decide the locations for future measurements. However, computing MI for reasonably-sized maps is slow and power hungry, which has been a bottleneck towards fast and efficient robotic exploration. In this paper, we introduce a new hardware accelerator architecture for MI computation that features a low-latency, energy-efficient MI compute core and an optimized memory subsystem that provides sufficient bandwidth to keep the cores fully utilized. The core employs interleaving to counter the recursive algorithm, and workload balancing and numerical approximations to reduce latency and energy consumption. We demonstrate this optimized architecture with a Field-Programmable Gate Array (FPGA) implementation, which can compute MI for all cells in an entire 201-by-201 occupancy grid ({\em e.g.}, representing a 20.1m-by-20.1m map at 0.1m resolution) in 1.55 ms while consuming 1.7 mJ of energy, thus finally rendering MI computation for the whole map real time and at a fraction of the energy cost of traditional compute platforms. For comparison, this particular FPGA implementation running on the Xilinx Zynq-7000 platform is two orders of magnitude faster and consumes three orders of magnitude less energy per MI map compute, when compared to a baseline GPU implementation running on an NVIDIA GeForce GTX 980 platform. The improvements are more pronounced when compared to CPU implementations of equivalent algorithms. 
\end{abstract}

\section{Introduction} \label{sec:introduction}

\begin{figure}[!ht]
\centerline{\includegraphics[width=0.85\linewidth]{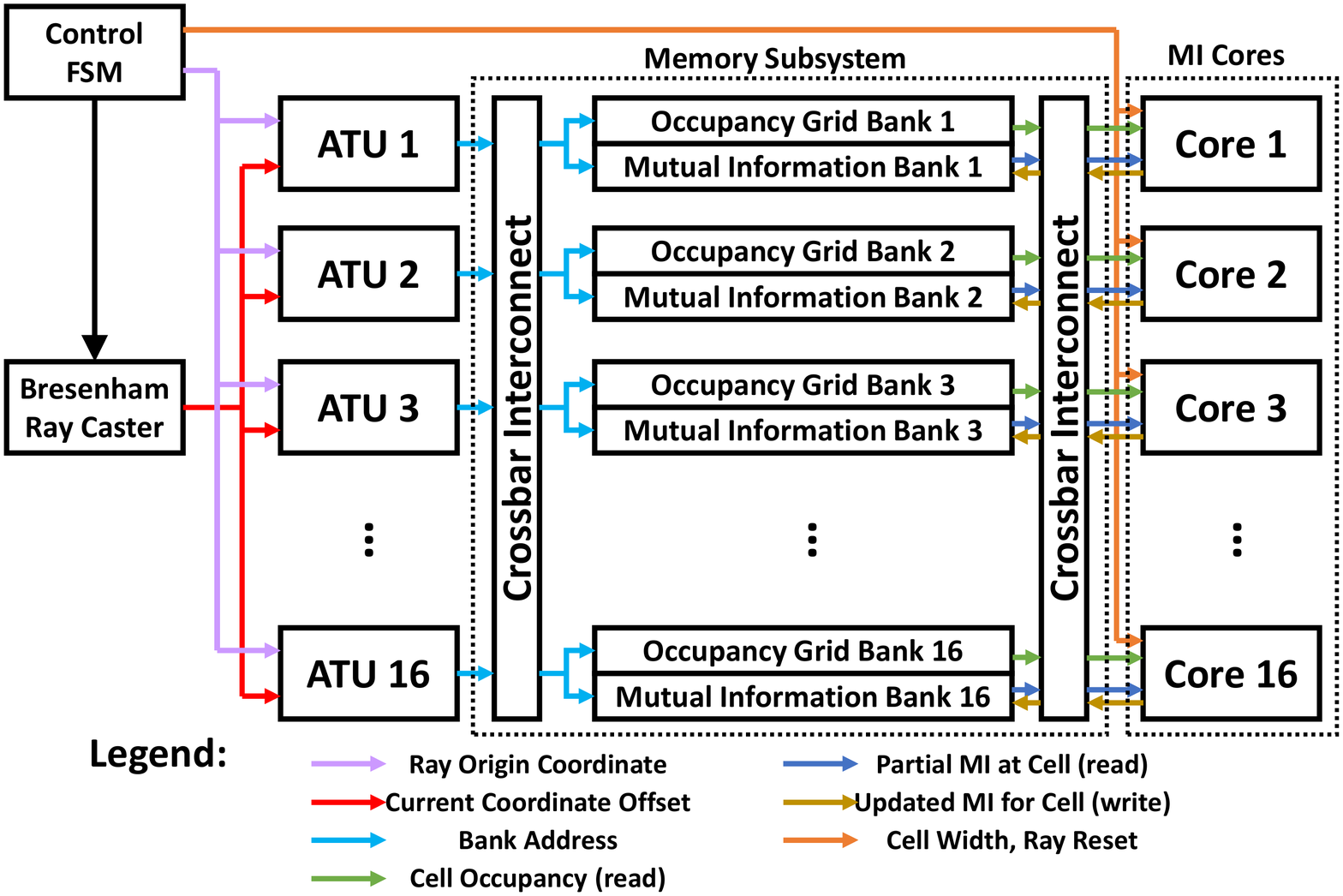}}
\centerline{\includegraphics[width=0.85\linewidth]{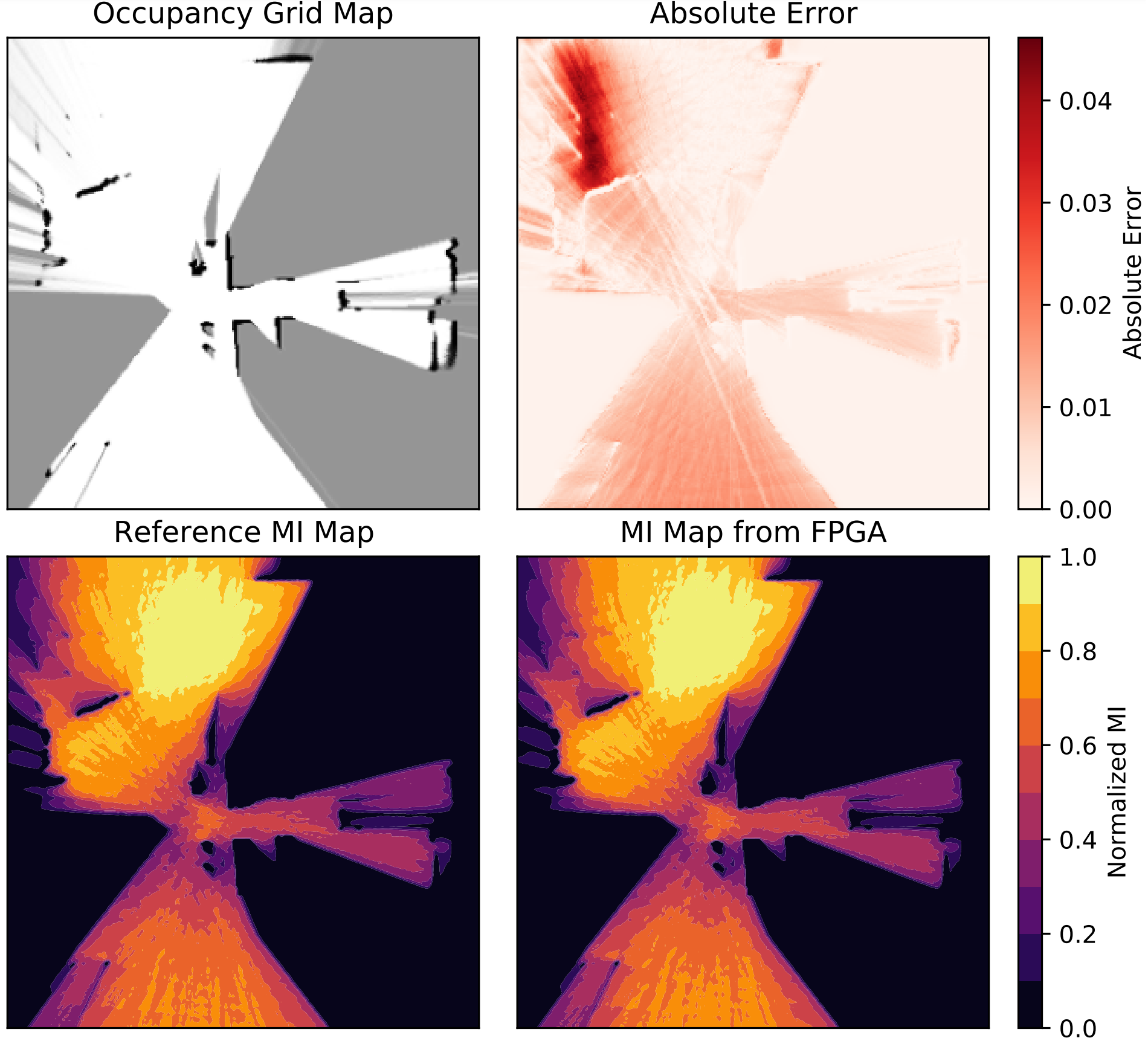}}
\caption{(top) Top-level hardware architecture comprised of Address Translation Units (ATU), crossbar interconnects, occupancy grid map, MI map, and $16$ FCMI computation cores. (bottom) An example input occupancy grid, corresponding MI map contours generated from a reference C++ implementation and our FPGA implementation, and the absolute error between the two.}
\label{fig:overall_arch_accuracy}
\end{figure}

Exploration problems arise in an increasing set of applications, particularly as robotic vehicles are being utilized in deep sea and space exploration missions as well as search and rescue scenarios. In most applications, the goal is to create a map of the environment, {\em e.g.}, from range measurements, as quickly as possible. A key problem, then, is to find the best locations for future measurements in order to achieve this goal. 


Consider an instance of this problem involving a robot equipped with a laser scanner. A typical computation cycle in this process involves making a scan of the robot's surroundings, updating the partially-unknown map with the information from the scan, determining the best location for the next scan and moving to that location. An information-theoretic approach to determining the best scan locations involves finding the location with the highest Mutual Information (MI) between the potential future measurement and the existing map. Unfortunately, computing MI at all map locations is slow and power hungry, even for small-size maps \cite{Cadena16tro-SLAMfuture}, which makes it especially infeasible for usage in medium to small scale robots with the energy budget of a few kiloJoules. Most heuristics used to reduce the scale of this computation, such as calculating the MI at sparse points instead of the entire map, fail to provide guarantees, {\em e.g.}, on the quality of the resulting motion~\cite{burgard2005coordinated, gonzalez2002navigation, holz2011comparative}. 







To accelerate the computation of MI, several novel algorithms were proposed. For instance, the efficient computation of Cauchy-Schwarz Quadratic Mutual Information (CSQMI) is introduced in~\cite{charrow2015csqmi, charrow2015information}. In addition, the high-throughput computation of Shannon Mutual information is proposed in the Fast Shannon Mutual Information (FSMI) algorithm~\cite{zhang2019fsmi}. More recently, the computation of Shannon MI is reformulated recursively and continuously along the range measurements in the Fast Continuous Mutual Information (FCMI) algorithm~\cite{henderson2020efficient}, which leads to a significant reduction in amount of compute needed for MI across an entire map.
Although these algorithms reduce the time complexity for MI computation, their optimized implementations on CPUs and GPUs still achieve a throughput that is orders-of-magnitude below what is required for autonomous exploration, especially in large environments. This motivates the design of specialized hardware accelerators for computing MI.

Hardware accelerators have been proposed for robotics applications such as motion-planning~\cite{atay2006motion, murray2016robot, kim2017brain, xiao2017parallel} and visual-inertial odometry~\cite{zhang2017visual, suleiman2018navion}. To our best knowledge, the first and only hardware accelerator for computing MI was proposed in \cite{peter2019rss}, which contains 16 high-throughput computation cores that compute FSMI. For moderately-sized environments (\emph{i.e.}, a map size of $200 \times 200$ to $512 \times 512$ cells), this accelerator computes MI for the entire map in \emph{near} real time (\emph{i.e.}, less than 10Hz), which is far below the rate in which the robot makes range measurements (around $30$-$60$Hz). 

The computation of mutual information may seem like a task that can be easily parallelized. One might be tempted to think, for example, that computational cores dedicated to various parts of the map can work in parallel to compute the mutual information for all cells in the map all at the same time. 
Indeed, under certain widely-utilized assumptions, the computation decouples and cells can be evaluated separately. 
Unfortunately, however, the computation does not scale, not because it can not be decoupled, but because it becomes impossible to feed all cores with the data that they need for their computations. More specifically, all cores need to access different portions of the map, which resides in memory, all at the same time. The memory bandwidth does not scale to keep all cores busy. In fact, we show in this paper that naively scaling up the number cores does not lead to a fast computation of the mutual information metric, as most cores end up underutilized due to insufficient memory bandwidth.

The main contribution of this paper is a scalable, multi-core hardware architecture for computing mutual information for a 2D occupancy grid map.
%
%
This architecture has two key components: (i) a memory management system that provides sufficient bandwidth to all cores, and (ii) a low-latency, energy-efficient core (\emph{i.e.} processing element) design that fully utilizes all available bandwidth.
Specifically, the memory management system exploits the memory access pattern of the cores and splits the storage of the occupancy grid map into smaller memory banks so that all cores access different banks every cycle.
%
%
The core contains a feedback architecture designed to handle the dependencies that are introduced by the recursive nature of the FCMI algorithm~\cite{henderson2020efficient}, so that all compute units within the core are fully utilized.  Each core is time-interleaved to compute the FCMI along multiple rays in order to avoid stalls associated with the feedback architecture.

The proposed hardware architecture is implemented with 16 cores on a Xilinx Zynq-7000 XC7Z045 FPGA. For a robot equipped with a 60-ray range sensor utilizing an occupancy grid map of size $201\times 201$, our FPGA implementation computes MI for the entire map in 1.55 ms (647 Hz, or much faster than real time), which is $381\times$ faster than an optimized C++ implementation running on a ARM Cortex A57 (embedded cellphone processor) and $71\times$ faster than an optimized CUDA implementation running on a NVIDIA GTX 980 GPU. In addition, our FPGA implementation consumes $1.7$ mJ of energy per MI map compute, which is $1314\times$ lower than ARM Cortex A57 and $2650\times$ lower than NVIDIA GTX 980 GPU. A CPU on a robot can therefore offload Mutual Information computation to such an accelerator, while interfacing with the sensors and actuators and performing motion planning. The extremely low latency and energy consumption of our accelerator allows robots to explore efficiently without relying on heuristic approximations even with the energy budget of a few Joules.


\section{Preliminaries} \label{sec:preliminaries}

\subsection{The FCMI Algorithm} \label{subsec:fcmi}
Occupancy grid map~\cite{elfes1989using} is a common probabilistic representation of the 2D environment and used in several well known autonomous exploration systems, such as the frontier-exploration algorithm~\cite{yamauchi1997frontier} and Shannon MI-based mapping algorithms~\cite{julian2014mutual, bourgault2002information, hoffmann2010mobile, kollar2008efficient, stachniss2005information}. In an information theoretic framework, MI is used to quantify the expected information gain from potential range measurements in an unknown environment. The unknown environment is represented by an occupancy grid map, which is updated continuously by the robot throughout an exploration trial. The FCMI algorithm~\cite{henderson2020efficient} computes the mutual information between the map and potential range measurement rays emanating from the sensors (\emph{e.g.}, LiDAR) on such robot.

Let $Z = \{Z_{\theta_1}, Z_{\theta_2} \ldots\}$ represent range measurements made by sensor rays with angles $\Theta = \{\theta_1,\theta_2,\ldots\}$ emanating from robot's location $x$. Let $\Delta_\theta$ denote the angular resolution of the sensor rays, which is constant across all rays. The Shannon MI $I_x(M; Z)$ between the measurements $Z$ made at location $x$ and map $M$ is defined as
\begin{equation} \label{eqn:mi}
    I_x(M; Z) = \sum_{\theta \in \Theta} I_x(M;Z_{\theta})\Delta_\theta.
\end{equation}

Note that the MI between a single range measurement $Z_\theta$ and map $M$ can be written as
\begin{equation} \label{eqn:mi_x}
    I_x(M; Z_{\theta}) = h_x(Z_{\theta}) - h_x(Z_{\theta} | M),
\end{equation}
where $h_x(Z_{\theta})$ is the entropy of the range measurement, and $h_x(Z_{\theta} | M)$ is the conditional entropy of the range measurement given the map.

Let $S$ denote the coordinates of cells in the occupancy grid map that are intersected by a sensor ray of infinite length with angle $\theta$. Suppose that the same sensor ray passes through location $x$ such that $x \in S$. For the cell at location $x$, let $o$ denote its occupancy probability, which is derived from previous range measurements. We define four expectations $\alpha_1, \beta_1, \alpha_0$ and $\beta_0$ (as define in~\cite{henderson2020efficient}) associated with the same cell as follows:
\begin{subequations} \label{eqn:expectations}
\begin{align}
    \alpha_1 &= e^{-\lambda_m w}\left((\alpha'_1 + \lambda w\beta'_1) + w(\alpha'_0 + \lambda_m w\beta'_0)\right),\nonumber\\
    &+ \lambda_m^{-1}\left(\gamma_3(\lambda_m w) - \gamma_2(\lambda_m w)\log\lambda_m\right),\\
    \beta_1 &= e^{-\lambda_m w}\left(\beta'_1 + w\beta'_0\right) + \lambda_m^{-1}\gamma_2(\lambda_m w),\\
    \alpha_0 &= e^{-\lambda_m w}\left(\alpha'_0 + \lambda_m w\beta'_0\right) + \gamma_2(\lambda_m w)\nonumber\\
    &- \gamma_1(\lambda_m w)\log\lambda_m,\\
    \beta_0 &= e^{-\lambda_m w} \beta'_0 + \gamma_1(\lambda_m w),    
\end{align}
\end{subequations}
where $\alpha'_1, \alpha'_0, \beta'_1, \beta'_0$ denote the four expectations from the previous cell along the same sensor ray, $w$ is the width of the current cell, $\Lambda$ is a large quantity used to barely distort the scan measurements ($\Lambda = 10^7$ for this work), $\lambda = -\log(1-o)$, $\lambda_m = \min(\lambda,\Lambda)$ and $\gamma_{s \in \{1,2,3\}}(\cdot)$ is the lower incomplete gamma function.

The entropy and conditional entropy in Eqn.~\eqref{eqn:mi_x} are approximated from the aforementioned expectations as follows:
\begin{subequations} \label{eqn:entropies}
\begin{align}
    h_x(Z_{\theta}) &\approx \alpha_1 \Delta_\theta, \\
    h_x(Z_{\theta}|M) &\approx (1 - \log \Lambda)\beta_1 \Delta_\theta.
\end{align}
\end{subequations}

Due to the recursive computation of the four expectations along a sensor ray, the entropy defined in Eqn.~\eqref{eqn:entropies} for every cell within a map of size $H \times H$ can be computed with a time complexity of $O(H^2)$. Let $|\Theta|$ denote the number of sensor rays emanating from the robot. Then, the mutual information defined in Eqn.~\eqref{eqn:mi} for the entire map can be computed with a time complexity of $O(|\Theta|H^2)$.


\subsection{A Lower Bound for Latency} \label{subsec:ideal_latency}
Let $n$ denote the number of cores in the hardware accelerator, which is clocked from a single source with frequency $F$. Suppose that the robot makes range measurements using $|\Theta|$ rays within an occupancy grid map of size $H \times H$. Recall that the FCMI algorithm has a time complexity of $O(|\Theta|H^2)$ for computing the corresponding MI map of the same size. If the hardware needs at least one cycle to compute the entropy defined in Eqn.~\eqref{eqn:entropies} for each cell along every sensor ray, the time between receiving an occupancy grid and producing the resultant MI (the lower bound latency of the system) can be determined from:
\begin{equation}\label{eqn:ideal_latency}
    t_\text{LB} = \frac{H^2|\Theta|}{nF},
\end{equation}
which has a unit of seconds, assuming unlimited memory bandwidth. 

In this work, we present a hardware architecture with $n=16$ cores clocked at $F = 100$ MHz. In \refsec{results}, we discuss how the achieved latency of the proposed architecture scales with the number of cores $n$, and compare it with the aforementioned lower bound.

\section{Hardware Architecture} \label{sec:architecture}


\subsection{Architecture Overview}

The proposed hardware architecture for FCMI computation is shown in \reffig{overall_arch_accuracy} (top) for a 16-core implementation. The architecture contains a memory subsystem that splits the occupancy grid map and MI map into 16 memory banks. The Bresenham ray caster determines the cells that are intersected by each ray. The Address Translation Unit (ATU) determines the memory storage location for each cell in the memory subsystem. There are two crossbar interconnects that function as switches for controlling the flow of data. The first crossbar connects the ATUs to the memory subsystem for memory request generation. The second crossbar connects the memory subsystem to the cores for data delivery and MI map update. To achieve low-latency computation, we need to ensure that (i) the memory subsystem can generate enough bandwidth to keep all cores active, and (ii) the cores can effectively utilize all available bandwidth without stalling.


\subsection{Memory Subsystem} \label{subsec:memory_subsystem}

\begin{figure*}
     \centering
     \begin{subfigure}[b]{0.23\textwidth}
         \centering
         \includegraphics[width=\textwidth]{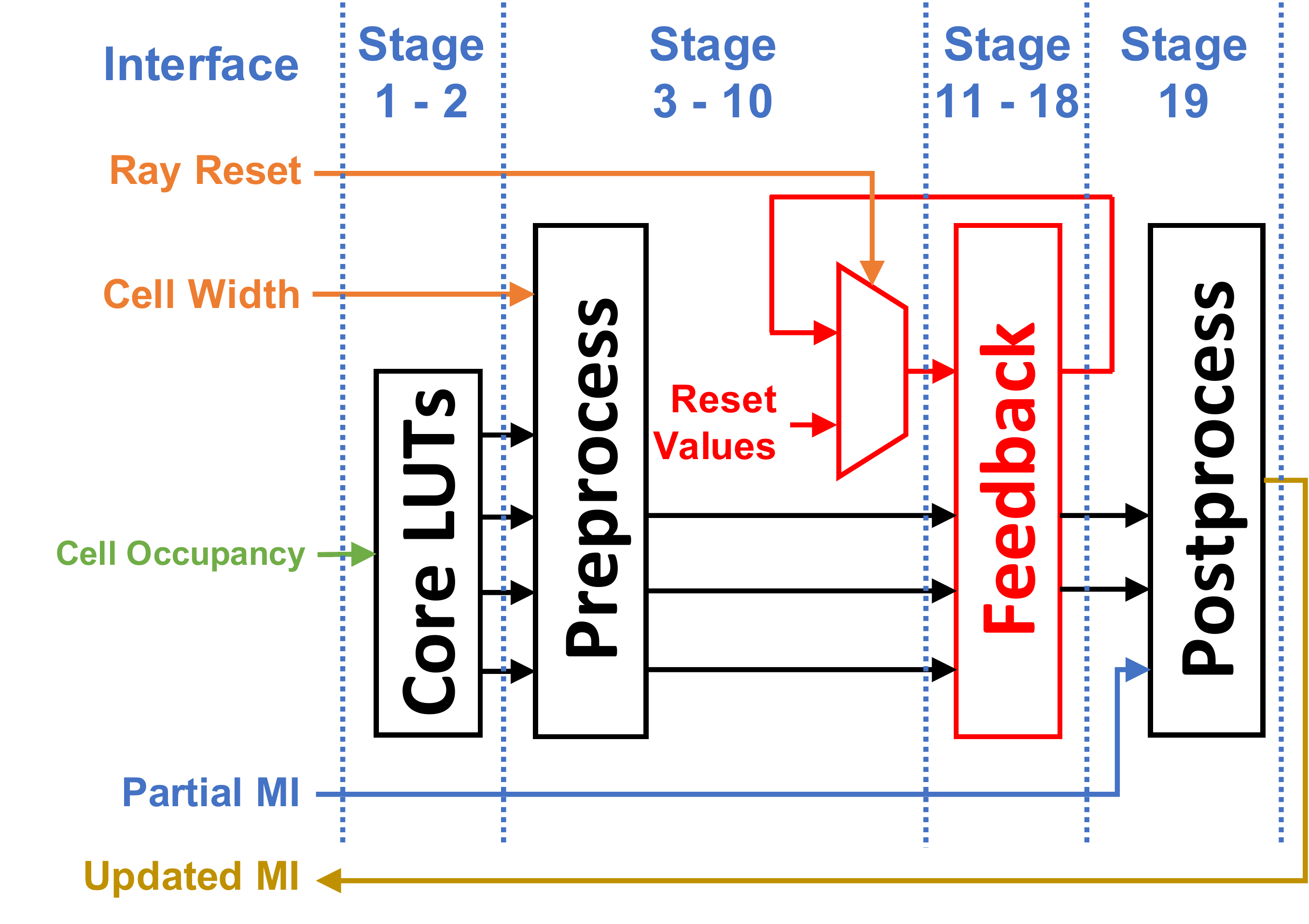}
         \caption{}
         \label{fig:core_arch}
     \end{subfigure}
     \hfill
     \begin{subfigure}[b]{0.17\textwidth}
         \centering
         \includegraphics[width=\textwidth]{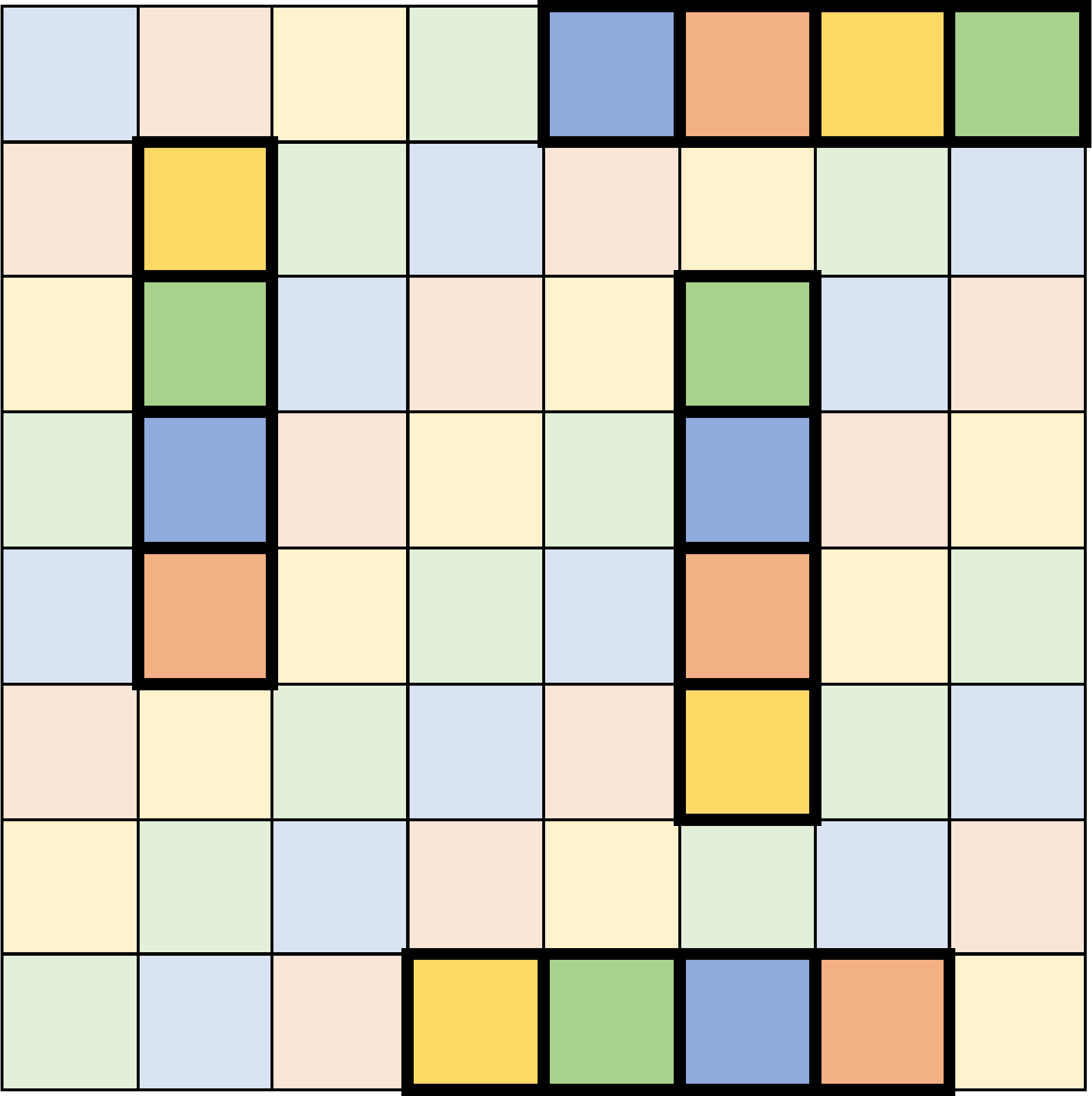}
         \caption{}
         \label{fig:banking}
     \end{subfigure}
     \hfill
     \begin{subfigure}[b]{0.42\textwidth}
         \centering
         \includegraphics[width=\textwidth]{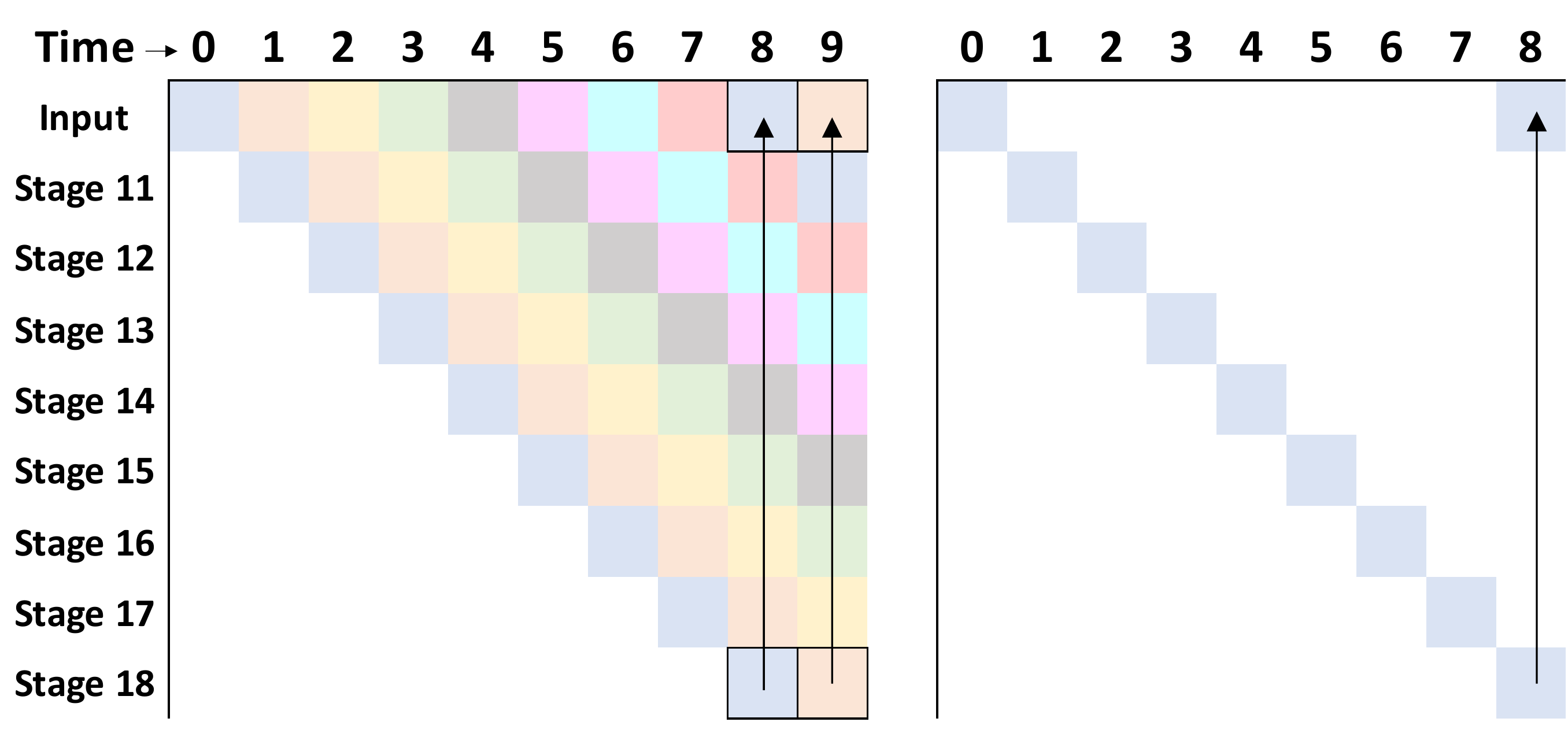}
         \caption{}
         \label{fig:interleaving}
     \end{subfigure}
        \caption{(a) The micro-architecture of the FCMI core comprised of three sections in the pipeline. (b) An example of the diagonal banking pattern with 4 banks, each in a different color. Any 4 consecutive cells here (examples highlighted) are guaranteed to lie on distinct banks. (c) Time is represented on the x axis and pipeline stages are represented on the y axis. Note that the feedback section occupies Stages 11-18 of the pipeline. (left) Interleaving 8 rays (represented with distinct colors) into a core. (right) Computing only one ray at a time, leading to a 7 cycle stall when waiting for results at the output stage.}
\end{figure*}


On the FPGA, memory units are implemented as dual-port SRAMs, which allows for two memory accesses in each cycle. Since the MI map is updated every cycle using one write port, only one read port from each SRAM is available for delivering data to the cores. If multiple cores attempt to concurrently access different cells stored within the same bank, a memory access conflict occurs and some cores must stall.

\begin{figure}[t]
\centerline{\includegraphics[width=0.9\linewidth]{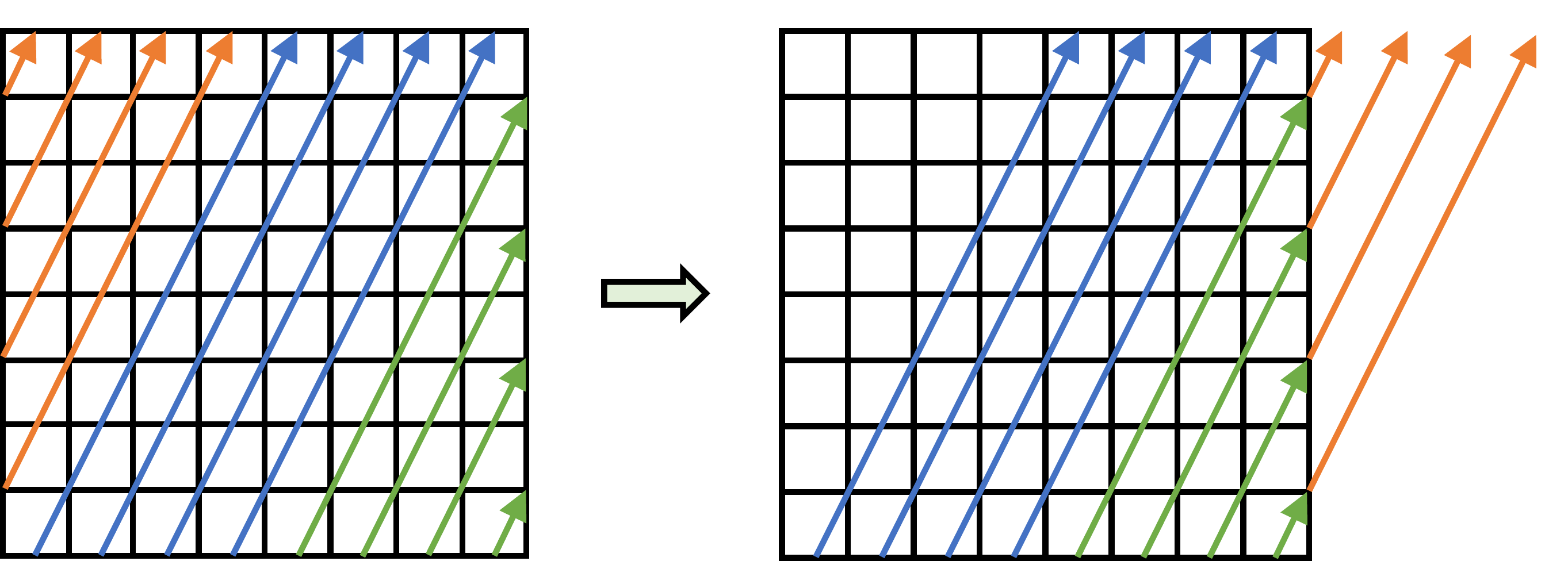}}
\caption{Wrapping rays around map boundaries to balance core workload and minimize latency.}
\label{fig:workload_balancing}
\end{figure}

Recall that the coordinates for each cell are computed using the Bresenham ray casting algorithm~\cite{Bresenham1965AlgorithmFC}, which we chose for its efficient hardware implementations~\cite{bresenham_hw}. Since all 16 cores concurrently compute FCMI along rays of the same angle as shown in Fig.~\ref{fig:workload_balancing}, the cores are guaranteed to access consecutive cells along a row or column of the map every cycle. This memory access pattern is inherent to the algorithm and is highly predictable. In order to exploit this property, the maps are partitioned and stored in 16 memory banks using a diagonal banking pattern, which ensures that any 16 consecutive cells within every row or column of the map are stored in distinct banks~\cite{peter2019rss}. An example of the diagonal banking pattern with 4 bank partitions is shown in \reffig{banking}, which can be extended to 16 banks. The occupancy grid map memory and the MI memory are both banked under this scheme.

\subsection{FCMI Core} \label{subsec:mi_core}
The micro-architecture of the FCMI core is illustrated in Fig.~\ref{fig:core_arch}. The cell accepts a new occupancy value and a partial MI for one cell along the sensor ray every cycle. The updated MI at the same cell is computed using the FCMI formulation and written back to the MI map in the memory subsystem. The low-latency, energy-efficient FCMI computation of each core is enabled by the following:

\subsection*{1) Arithmetic Optimizations} \label{subsec:arithmetic_opt} 
We store and compute MI results using 32-bit fixed point numbers (20 integer, 12 fractional bits), since adders and multipliers implemented using fixed point numbers tend to be faster and more energy efficient using the specialized DSPs on the FPGA. The choice of 32-bit width comes from taking into account the dynamic range of numbers needed to avoid under/over-flow in this application. In addition, the core needs to evaluate the following logarithms and reciprocals: $v = 1 - o$, $\lambda = -\log v$, $-\log\lambda$, and $1/\lambda$ (from Eqn.~\eqref{eqn:expectations}). They are all direct functions of the occupancy probabilities $o$ which is uniformly quantized into $101$ values such that $o \in \{0,0.1,\ldots,1\}$. Note that $o = 0$ and $o = 1$ are degenerate cases, which are handled using pre-calculated results. A small look-up table is used which store all values of logarithms and reciprocals. This look-up table occupies only two Block RAM tiles per core on the FPGA.

Finally, the core also needs to evaluate an exponential to compute $e^{-\lambda w}$, where $\lambda$ has one-to-one correspondence to the occupancy probability $o$ (101 values) and $w$ is the cell width. To the best of our knowledge, no such module exists that uses fixed point arithmetic that would lead to a throughput of one result per clock cycle. To achieve such a throughput, a look-up table or an approximation should be used. Note that $w$ is dependent on number of sensor rays $|\Theta|$ such that $w = \left\lceil|\Theta|/8\right\rceil$ (accounting for symmetries w.r.t. the axes).
If $|\Theta|=60$, a look-up table mapping all possible combinations of $(\lambda, w)$ to the corresponding $e^{-\lambda w}$ value would need to contain $\lceil101(60)/8\rceil = 808$ values, which translates to $16$ Block RAM tiles per core on the target FPGA. The FPGA does not have sufficient Block RAM to support the look-up table for $16$ cores. In addition, since the size of the look-up table is dependent on the angular resolution of the sensor rays, the look-up table implementation would only be able to support a limited set of sensor configurations. From formulation of FCMI, all $-\lambda w$ values lie inside $[-8, 0]$. Thus, we use a piece-wise linear approximation to the exponential function that splits the range into 16 uniform pieces and uses a mean-squared error minimizing linear approximation to each. Such approximation achieves a maximum relative error of $\pm 3\%$, while not imposing a limit on sensor configurations, and consuming only one Block RAM per core.

\subsection*{2) Hardware Pipeline} 
As evident from Eqn.~\eqref{eqn:expectations}, each cell update is a long chain of mathematical operations. In order to reduce the critical path of the system to allow a faster clock, we pipeline these operations into 19 stages and perform the operations over consecutive clock cycles. The pipeline allows the system to be clocked at $100 \text{ MHz}$ instead of the combined combinational delay of the system, which would be close to only $5 \text{ MHz}$.

Since the quantities in Eqn.~\eqref{eqn:expectations} are defined recursively, the pipeline contains a feedback loop which increases the latency of the computation. A na\"ive implementation of the pipeline executes the operations in Eqn.~\eqref{eqn:expectations} for each cell in 19 clock cycles such that each pipeline stage stalls for 18 cycles. As a solution, the 18 stages of the pipeline were split into three sections: Preprocess, Feedback and Postprocess (as shown in Fig.~\ref{fig:core_arch}). This reduces the feedback loop to only 8 stages in the Feedback section. In addition, we introduce ray-interleaving to eliminate the remaining stalls associated with the smaller feedback loop in \refsubsec{ray_interleaving_balancing}.

\subsubsection*{Preprocess}
This stage uses the cell occupancy value to compute as many intermediate values as possible without requiring information about the last cell. The look-up tables are accessed to obtain $\lambda$, $-\log \lambda$ and $1/\lambda$ defined in Section~\ref{subsec:fcmi}. The exponential $e^{-\lambda w}$ and the lower incomplete gamma function values $\gamma_0(\lambda w)$, $\gamma_1(\lambda w)$, $\gamma_2(\lambda w)$ are also precalculated for use in the feedback stage.

\subsubsection*{Feedback}
This stage uses the intermediate values from the Preprocess stage and the values of $\{\alpha'_1, \beta'_1, \alpha'_0, \beta'_0\}$ for the previous cell along the sensor ray to compute the updated $\{\alpha_1, \beta_1, \alpha_0, \beta_0\}$ values for the current cell.

\subsubsection*{Postprocess}
This stage uses the updated $\{\alpha_1, \beta_1, \alpha_0, \beta_0\}$ values for the current cell in addition to the current partial MI from the memory subsystem to compute the updated MI of the same cell.

\subsection*{3) Ray Interleaving \& Wrapping} \label{subsec:ray_interleaving_balancing}
The process of interleaving is represented visually in \reffig{interleaving}. From the same figure, the recursion for the next cell along the sensor ray begins after the previous cell along the same ray arrives at the final stage of the Feedback section (stage 18). In the absence of ray interleaving, the core needs to stall for 8 cycles for every cell along the sensor ray, which leads to an increase in latency.

Due to ray interleaving, each core interleaves the computation of FCMI for 8 rays simultaneously to achieve a throughput of 1 cell per cycle. Therefore, the amount of time a core spends processing each ray is proportional to its length. In addition to ray interleaving, the workload is balanced across all the cores that process rays with different lengths via wrapping. An example is shown in \reffig{workload_balancing} where the four cores that concurrently process the green rays complete at different times. Note that orange rays wrap around the green rays across the boundary of the map. By assigning the computation of the orange rays to the same four cores after the green rays, the workload across all cores becomes balanced.


\section{Results}~\label{sec:results}
The proposed architecture is implemented on a Xilinx Zynq-7000 XC7Z045 FPGA which is clocked at 100MHz and can store a maximum map size of $512 \times 512$.
We also evaluate the latency, energy per MI map compute and scalability of the proposed architecture with respect to the number of cores. Finally, we present the impacts that the proposed architecture can have in an exploration experiment.

\subsection{Accuracy} \label{subsec:accuracy}

To evaluate the accuracy of our implementation, we generated contours lines from MI maps for a variety of occupancy grid maps from real world exploration trials.

In Fig.~\ref{fig:overall_arch_accuracy} (bottom), we compare the locations of high MI generated from our FPGA implementation with those from a reference C++ implementation. The MI map normalized to $[0, 1]$ computed by the FPGA differs only marginally from the one generated from the reference C++ implementation (by less than $0.05$ across all points), the major source of error being the piece-wise linear approximation of the exponential
function (Section~\ref{subsec:arithmetic_opt}). We see that our generated MI map is sufficiently accurate for real world exploration trials, as seen from Section~\ref{subsec:entropy}.

\subsection{Latency} \label{subsec:latency}
We evaluate the latency (\emph{i.e.} the time elapsed between receiving an occupancy grid and producing the corresponding MI map) of our FPGA implementation that computes MI for an entire map of size $201 \times 201$ cells with $60$ sensor rays. This map captures a robot in a $20.1 \text{m } \times 20.1 \text{m}$ environment mapped at $10$ cm resolution. The FPGA computes MI for the entire map in 1.55 ms using 16 cores. 

Fig.~\ref{fig:latency_energy} illustrates that the latency of our FPGA implementation reaches within $2\%$ of the lower bound (Eqn.~\eqref{eqn:ideal_latency}) and scales linearly with the number of cores.
Without workload balancing or ray interleaving (Section~\ref{subsec:ray_interleaving_balancing}), the latency would increase by $22.9 \%$ or $8$ times respectively. Finally, if the occupancy grid map and MI map are stored in one bank (\emph{i.e.} no banking), the latency does not decrease beyond what is achievable with a single core because the memory cannot deliver enough bandwidth.

We compared the latency of our FPGA implementation that accelerates the FCMI algorithm with the optimized C++ implementation on ARM Cortex A57 and Intel Core i7-1065G7 (embedded mobile and desktop processors, respectively), and optimized CUDA implementation on NVIDIA GTX 980 GPU. In addition, we compared our FPGA implementation that accelerates the FCMI algorithm with a prior FPGA implementation \cite{peter2019rss} that accelerates the FSMI algorithm. 
The results are shown in \reffig{latency_energy}. Our FPGA implementation is $71 \times$ faster than the GTX 980, $104 \times$ faster than the FPGA implementation for FSMI, $190 \times$ faster than the Core i7 CPU, and $381 \times$ faster than the Cortex A57 CPU.

The system latency scales in proportion to the number of pixels in the map $(H^2)$, and the number of sensor rays $(|\Theta|)$. For example, the system computes the MI for a $256\times256$ map in 2.51 ms, and for a $512\times512$ map in 10.1 ms, and the relative speed-up over the baseline is maintained.

\subsection{Energy Consumption} \label{subsec:energy}
\begin{figure}
\centering
\includegraphics[width=0.9\linewidth]{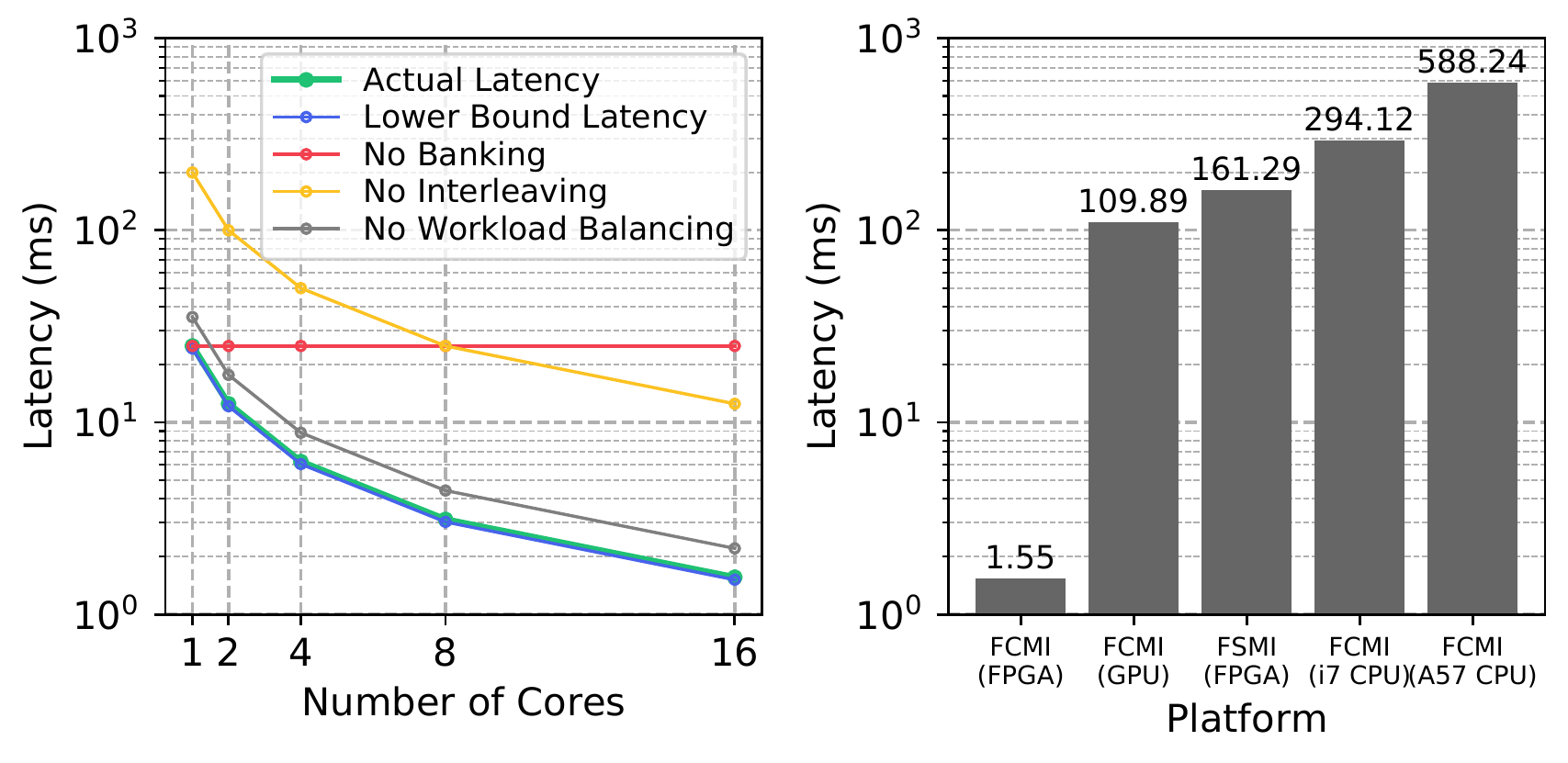}
\includegraphics[width=0.9\linewidth]{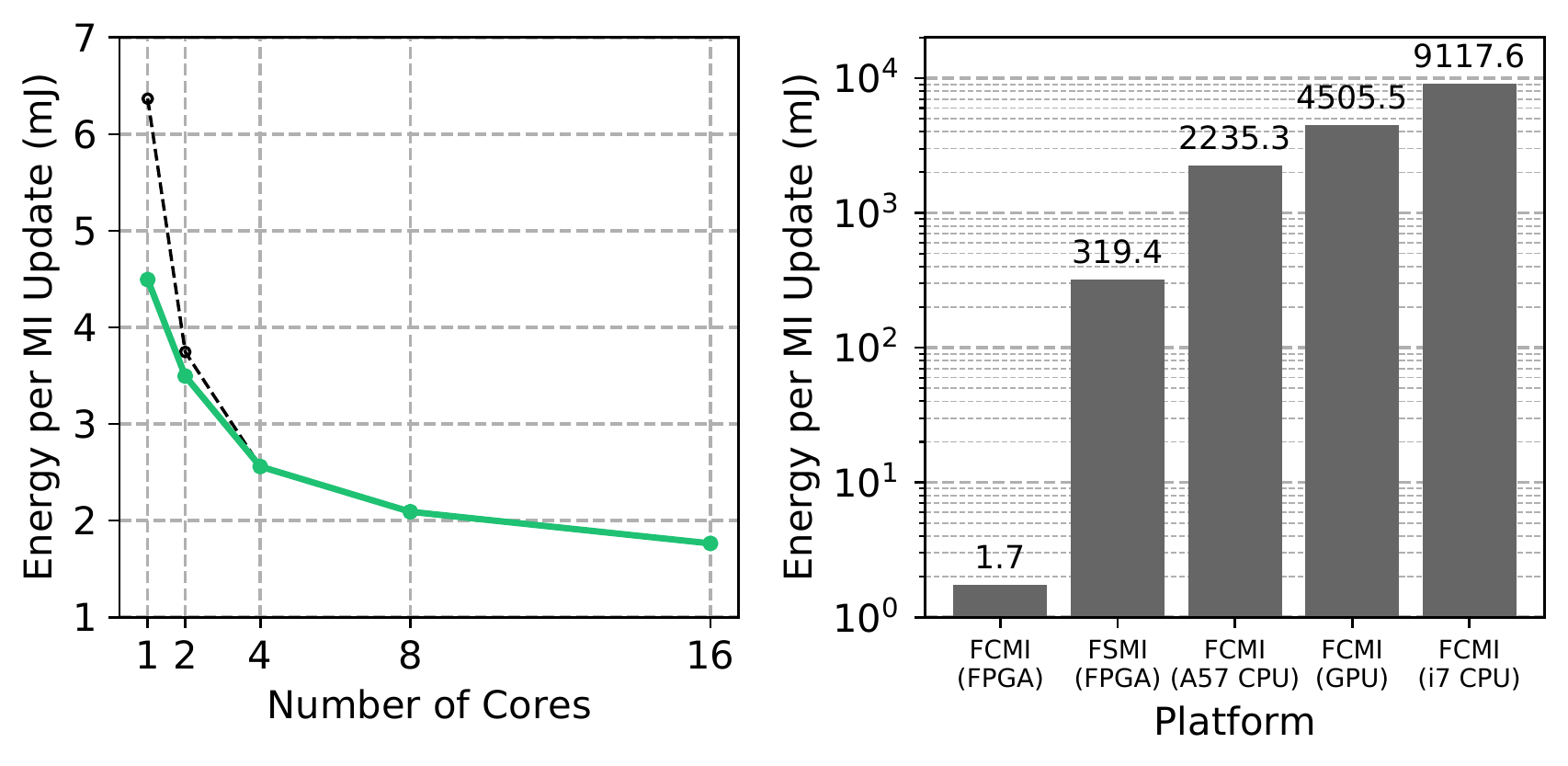}
\caption{(top, left) Impact of the number of cores on system latency. (top, right) A comparison of latency achieved across platforms. (bottom, left) Impact of the number of cores on energy consumed per MI compute. \textit{[Errata: Careful re-simulation shows updated (better) numbers for Energy per MI Update. The numbers from the original publication are shown by the black dashed line.]} (bottom, right) A comparison of the energy consumption per MI map compute across platforms. FSMI (FPGA) is the previous work~\cite{peter2019rss} and FCMI (FPGA) is this work. Both architectures contain 16 cores and are implemented on the same FPGA.}
\label{fig:latency_energy}
\end{figure}

We evaluate the energy per MI map compute (\emph{i.e.}, the energy consumed by the FPGA for every computation of an MI map from an occupancy grid) on the same map as Section~\ref{subsec:latency}. \reffig{latency_energy} compares this metric of our FPGA implementation with other hardware platforms, and illustrates how it scales with the number of cores. The energy consumption is obtained post-place-and-route with annotated switching activity.

With $16$ cores clocked at $100$ MHz, our FPGA implementation consumes only $1.7$ mJ of energy per MI map compute, which is about $2650 \times$ lower than the GTX 980, $5360 \times$ lower than the Core i7 CPU, $1314 \times$ lower than the Cortex A57 CPU and $187 \times$ lower than the FPGA implementation for FSMI.

\subsection{Resource Utilization}

\begin{table}[b]
\caption{Resource utilization of the proposed architecture with 16 cores on a Xilinx Zynq-7000 XC7Z045 FPGA that stores a maximum map size of $512 \times 512$.}
\begin{center}
\begin{tabular}{|c||c|c|c|c|c|}
\hline
\multirow{2}{*}{\textbf{Module}} & \textbf{LUT} & \textbf{LUT} & \multirow{2}{*}{\textbf{FF}} & \multirow{2}{*}{\textbf{BRAM}} & \multirow{2}{*}{\textbf{DSP}}\\
& (as Logic) & (as RAM) & & & \\
\hline
\hline
Ray Caster & 1258 & 0 & 56 & 0 & 0\\
\hline
MI Memory & 10979 & 7 & 1123 & 232 & 0 \\
\hline
Occ. Memory & 2407 & 0 & 282 & 56 & 0 \\
\hline
Cores (16x) & 2970 & 170 & 2339 & 2 & 45 \\
\hline
\hline
\textbf{Total} & \textbf{61729} & \textbf{2699} & \textbf{38991} & \textbf{320} & \textbf{720} \\
\hline
\textbf{Total\%} & \textbf{29\%} & \textbf{4\%} & \textbf{9\%} & \textbf{59\%} & \textbf{80\%} \\
\hline
\end{tabular}
\label{tab:resource}
\end{center}
\end{table}

\reftab{resource} shows the resource utilization breakdown of our FPGA implementation with 16 cores. The utilization of the entire system scales linearly with respect to the number of cores, which offers great flexibility when implemented on systems with different computational resources and energy consumption. Although the size of the crossbar interconnect scales quadratically with the number of cores, it occupies a very small fraction of the design and this growth does not pose a problem for up to 16 cores.

\subsection{Impact on Exploration} \label{subsec:entropy}

\begin{figure}[t]
\centerline{\includegraphics[width=0.9\linewidth]{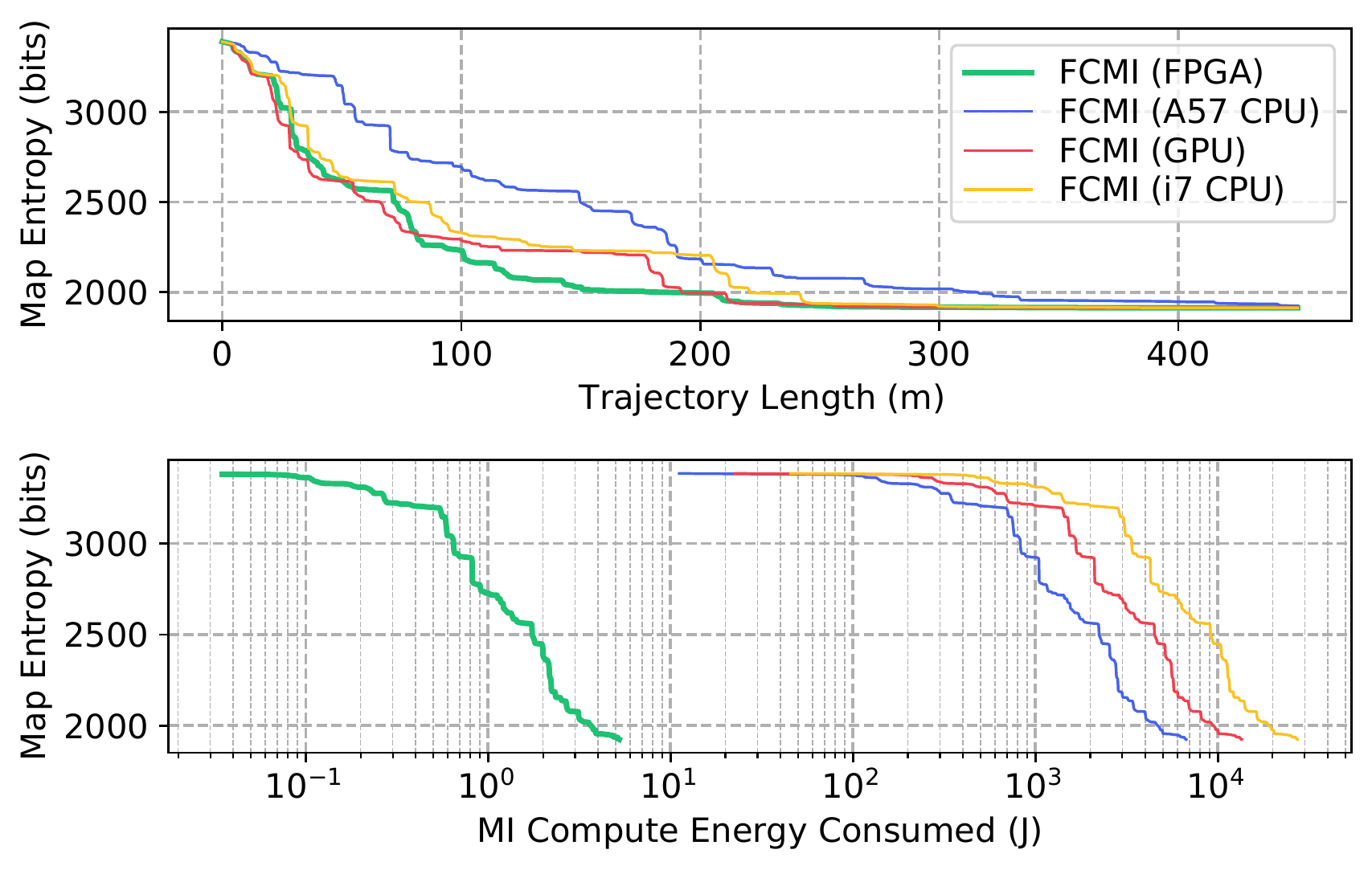}}
\caption{(top) Map entropy of an example map as a function of trajectory length using GBL exploration \cite{gblExplore} with MI computed at the maximum rates achievable on several platforms. (bottom) Map entropy on the same map as a function of total energy spent on MI compute with MI being computed at the rate matching the Cortex A57 CPU on all platforms in order to highlight the energy efficiency of our system.}
\label{fig:entropy_trajectory}
\end{figure}

We simulate a model robot using the Gonz\'alez-Ba\~nos-Latombe exploration strategy~\cite{gblExplore} to explore a sample map sized at $201 \times 201$ cells. The robot makes a LiDAR scan and updates its occupancy grid at 30 Hz \emph{i.e.} every 33 ms (a typical LiDAR sensor rate). Our system finishes the MI compute in 1.55 ms and remains idle while waiting for an update to the occupancy grid. Since the other platforms require more than 33 ms to finish the MI compute, they compute on the latest available occupancy grid. \reffig{entropy_trajectory} (top) shows map entropy of an example map as a function of trajectory length. It is evident that being able to compute the MI map at a faster rate leads to a quicker drop in map entropy.

More importantly, we measure the MI compute energy consumption during exploration on several platforms while performing MI computes at the same rate (\emph{i.e.}, 1.7 Hz which matches the Cortex A57 CPU) in order to highlight the energy efficiency of our architecture. From \reffig{entropy_trajectory} (bottom), our system completes the exploration task while spending around 5 J of energy on MI compute, roughly three orders of magnitude lower than the compared platforms. For context on the scale of these numbers, a typical cellphone battery holds about 40 kJ of energy \cite{SamsungS9Specs}, which would drain after about 5-6 map exploration trials running on its embedded processor. Our system, in comparison, would be able to perform around 8,000 such trials on that energy budget.

\section{Conclusions} \label{sec:conclusion}
In this work, we presented a novel scalable, multi-core architecture that exploits the recursive nature of the FCMI algorithm. The key contributions in this architecture are a memory subsystem that provides sufficient bandwidth to all cores, and a low-latency, energy-efficient core design that interleaves the computation across multiple sensor rays in order to utilize all available memory bandwidth.
We implemented the proposed architecture with 16 cores on an FPGA, which computes the \emph{entire} mutual information map for a given occupancy grid map in 1.55 ms (\emph{i.e.}, 647 Hz) while consuming only 1.7 mJ of energy per MI map compute. Not only does such a result lead to a significant reduction in time, trajectory and energy consumption during autonomous exploration, it also enables Mutual Information based exploration for medium and small scale robots for the first time. In addition, the latency and FPGA resource usage of the proposed architecture is scalable with the number of cores, which makes the architecture re-configurable for both high performance and lower energy applications, ranging from high-speed rovers to low-power micro-drones. 

{\bf Acknowledgements.}
This work was partially funded by the NSF RTML 1937501 and the NSF CPS 1837212.

\clearpage

\bibliographystyle{ieeetr}
\bibliography{references}

\begin{thebibliography}{10}

\bibitem{Cadena16tro-SLAMfuture}
C.~Cadena, L.~Carlone, H.~Carrillo, Y.~Latif, D.~Scaramuzza, J.~Neira, I.~Reid,
  and J.~Leonard, ``Past, present, and future of simultaneous localization and
  mapping: Towards the robust-perception age,'' {\em {IEEE Transactions on
  Robotics}}, vol.~32, no.~6, p.~1309–1332, 2016.

\bibitem{burgard2005coordinated}
W.~Burgard, M.~Moors, C.~Stachniss, and F.~E. Schneider, ``Coordinated
  multi-robot exploration,'' {\em IEEE Transactions on robotics}, vol.~21,
  no.~3, pp.~376--386, 2005.

\bibitem{gonzalez2002navigation}
H.~H. Gonz{\'a}lez-Banos and J.-C. Latombe, ``Navigation strategies for
  exploring indoor environments,'' {\em The International Journal of Robotics
  Research}, vol.~21, no.~10-11, pp.~829--848, 2002.

\bibitem{holz2011comparative}
D.~Holz, N.~Basilico, F.~Amigoni, S.~Behnke, {\em et~al.}, ``{A Comparative
  Evaluation of Exploration Strategies and Heuristics to Improve Them},'' in
  {\em ECMR}, pp.~25--30, 2011.

\bibitem{charrow2015csqmi}
B.~Charrow, S.~Liu, V.~Kumar, and N.~Michael, ``{Information-theoretic mapping
  using Cauchy-Schwarz Quadratic Mutual Information},'' in {\em IEEE
  International Conference on Robotics and Automation}, 2015.

\bibitem{charrow2015information}
B.~Charrow, G.~Kahn, S.~Patil, S.~Liu, K.~Goldberg, P.~Abbeel, N.~Michael, and
  V.~Kumar, ``{Information-Theoretic Planning with Trajectory Optimization for
  Dense 3D Mapping},'' in {\em Robotics: Science and Systems}, vol.~6, 2015.

\bibitem{zhang2019fsmi}
Z.~Zhang, T.~Henderson, V.~Sze, and S.~Karaman, ``{FSMI: Fast computation of
  Shannon Mutual Information for Information Theoretic Mapping},'' in {\em IEEE
  International Conference on Robotics and Automation}, 2019.

\bibitem{henderson2020efficient}
T.~Henderson, V.~Sze, and S.~Karaman, ``An efficient and continuous approach to
  information-theoretic exploration,'' in {\em 2020 IEEE International
  Conference on Robotics and Automation (ICRA)}, pp.~8566--8572, IEEE, 2020.

\bibitem{atay2006motion}
N.~Atay and B.~Bayazit, ``A motion planning processor on reconfigurable
  hardware,'' in {\em Robotics and Automation, 2006. ICRA 2006. Proceedings
  2006 IEEE International Conference on}, pp.~125--132, 2006.

\bibitem{murray2016robot}
S.~Murray, W.~Floyd-Jones, Y.~Qi, D.~J. Sorin, and G.~Konidaris, ``{Robot
  Motion Planning on a Chip},'' in {\em Robotics: Science and Systems}, 2016.

\bibitem{kim2017brain}
Y.~Kim, D.~Shin, J.~Lee, and H.-J. Yoo, ``{BRAIN: A Low-Power Deep Search
  Engine for Autonomous Robots},'' {\em IEEE Micro}, vol.~37, no.~5,
  pp.~11--19, 2017.

\bibitem{xiao2017parallel}
S.~Xiao, N.~Bergmann, and A.~Postula, ``{Parallel RRT∗ architecture design
  for motion planning},'' in {\em Field Programmable Logic and Applications
  (FPL), 2017 27th International Conference on}, pp.~1--4, 2017.

\bibitem{zhang2017visual}
Z.~Zhang, A.~A. Suleiman, L.~Carlone, V.~Sze, and S.~Karaman,
  ``{Visual-Inertial Odometry on Chip: An Algorithm-and-Hardware Co-design
  Approach},'' in {\em Robotics: Science and Systems}, 2017.

\bibitem{suleiman2018navion}
A.~Suleiman, Z.~Zhang, L.~Carlone, S.~Karaman, and V.~Sze, ``{Navion: A Fully
  Integrated Energy-Efficient Visual-Inertial Odometry Accelerator for
  Autonomous Navigation of Nano Drones},'' in {\em 2018 IEEE Symposium on VLSI
  Circuits}, pp.~133--134, 2018.

\bibitem{peter2019rss}
P.~Z.~X. Li, Z.~Zhang, S.~Karaman, and V.~Sze, ``{High-throughput Computation
  of Shannon Mutual Information on Chip},'' in {\em Robotics: Science and
  Systems (RSS)}, 2019.

\bibitem{elfes1989using}
A.~Elfes, ``Using occupancy grids for mobile robot perception and navigation,''
  {\em Computer}, no.~6, pp.~46--57, 1989.

\bibitem{yamauchi1997frontier}
B.~Yamauchi, ``A frontier-based approach for autonomous exploration,'' in {\em
  IEEE International Symposium on Computational Intelligence in Robotics and
  Automation}, pp.~146--151, 1997.

\bibitem{julian2014mutual}
B.~J. Julian, S.~Karaman, and D.~Rus, ``On mutual information-based control of
  range sensing robots for mapping applications,'' {\em The International
  Journal of Robotics Research}, vol.~33, no.~10, pp.~1375--1392, 2014.

\bibitem{bourgault2002information}
F.~Bourgault, A.~A. Makarenko, S.~B. Williams, B.~Grocholsky, and H.~F.
  Durrant-Whyte, ``Information based adaptive robotic exploration,'' in {\em
  Intelligent Robots and Systems, 2002. IEEE/RSJ International Conference on},
  vol.~1, pp.~540--545, IEEE, 2002.

\bibitem{hoffmann2010mobile}
G.~M. Hoffmann and C.~J. Tomlin, ``Mobile sensor network control using mutual
  information methods and particle filters,'' {\em IEEE Transactions on
  Automatic Control}, vol.~55, no.~1, pp.~32--47, 2010.

\bibitem{kollar2008efficient}
T.~Kollar and N.~Roy, ``{Efficient Optimization of Information-Theoretic
  Exploration in SLAM},'' in {\em AAAI}, vol.~8, pp.~1369--1375, 2008.

\bibitem{stachniss2005information}
C.~Stachniss, G.~Grisetti, and W.~Burgard, ``Information gain-based exploration
  using rao-blackwellized particle filters.,'' in {\em Robotics: Science and
  Systems}, vol.~2, pp.~65--72, 2005.

\bibitem{Bresenham1965AlgorithmFC}
J.~Bresenham, ``Algorithm for computer control of a digital plotter,'' {\em IBM
  Syst. J.}, vol.~4, pp.~25--30, 1965.

\bibitem{bresenham_hw}
S.~A. Edwards, ``{Bresenham’s Line Algorithm in Hardware},'' 2012.
\newblock URL:
  \url{http://www.cs.columbia.edu/~sedwards/classes/2012/4840/lines.pdf}.

\bibitem{gblExplore}
H.~H. González-Baños and J.-C. Latombe, ``Navigation strategies for exploring
  indoor environments,'' {\em The International Journal of Robotics Research},
  vol.~21, no.~10-11, pp.~829--848, 2002.

\bibitem{SamsungS9Specs}
``Samsung s9 technical specifications: Gsmarena.''
  \url{https://www.gsmarena.com/samsung_galaxy_s9-8966.php}.

\end{thebibliography}

\end{document}